\documentstyle[12pt,epsf]{article}
\begin{document}
\input epsf
\title{E-B Mixing in T-violating Superconductors}
\author{J. Goryo\thanks{e-mail:goryo@particle.sci.hokudai.ac.jp}
~and~K. Ishikawa\thanks{e-mail:ishikawa@particle.sci.hokudai.ac.jp} \\
{\it Department of Physics, Hokkaido University,}\\
{\it Sapporo, 060 Japan}}
\maketitle

\begin{abstract}
{ We analyze time-reversal-violating  processes  of  the p-wave superconductor.
The Landau-Ginzburg effective action has an induced T-violating term of 
electromagnetic  potentials which resembles the Chern-Simons term and 
causes  mixing between the electric  and  magnetic fields. Several 
T-violating electromagnetic phenomena caused by this term, such as an unusual
Meissner effect, the Hall effect without 
a magnetic field, and Faraday rotation 
without a magnetic field are investigated.} 
\end{abstract}

\newpage

 Time reversal symmetry holds in ordinary matter in  macroscopic systems 
because the electromagnetic interaction, which is  the origin of all 
forces in these regions, preserves the time-reversal symmetry. The symmetry is
broken explicitly by an external magnetic field or  spontaneously by  
condensation of a T-odd order parameter. Interesting 
physical phenomena would then occur. The Chern-Simons term is a T-odd bilinear
form of  electromagnetic potentials and has one  derivative . Hence 
this term is the lowest dimensional gauge invariant object and plays 
important roles in the low-energy and long-distance physics of the T-violating 
system\cite{2+1C-S1,2+1C-S2,2+1C-S3,2+1C-S4,2+1C-S5,2+1C-S6}.

 The Chern-Simons term plays an important role in the quantum Hall effect
\cite{2+1C-S5,2+1C-S6}\cite{QHE1,QHE2,QHE3,QHE4},where the time-reversal 
symmetry 
is broken 
explicitly by the external magnetic field.
In this paper we discuss time-reversal-violating processes of the p-wave 
superconductor\cite{Sup1}\cite{Sup2} in which time-reversal symmetry 
is broken spontaneously. We 
assume that pairs with the angular momentum of one along the z-axis condense   
homogeneously as  in the A-phase of $^{3}$He \cite{He}, and 
we also call this phase  as the A-phase of the p-wave superconductor
\cite{Sr2RuO41}\cite{Sr2RuO42}. 
We show, in this letter, several 
unusual electromagnetic phenomena, such as  mixing between 
the electric  and  magnetic fields (E-B mixing) 
in the superconductor.

Our discussion below may be valid for other T-violating superconductors, 
because the discussion does not depend on the detail forms of order parameters,
 but  only on the fact that Landau-Ginzburg action has a T-odd Chern-Simons- 
like term.

The long-range effective action for gauge fields in the A-phase of the $^{3}$He
 like superconductor at the Fermion 1-loop 
level is given by  \cite{Vol.1} \cite{Vol.2}\cite{GI}  
\begin{eqnarray}
  S_{{\rm eff}}^{({\rm f})}
& =\int d^{\rm D}x [-\frac{1}{4}F_{\mu\nu}F^{\mu\nu}
  + \frac{m_{\rm B}^{2}}{2}(\frac{1}{c^{2}}A_{0}^{2} - A_{i}^{2}) &\nonumber\\
& + \frac{\sigma}{2}e_{ij}(A_{0}\partial_{i}A_{j}+A_{i}\partial_{j}A_{0})] 
  + {\cal{O}}(e^{3}).&\label{10} 
\end{eqnarray}
 D represents  space-time dimension of the system and we consider cases of  
both 2+1 and 3+1 dimensions . 
The second term in  eq.(\ref{10}) represents the Meissner 
effect and $1/m_{\rm B}$ shows the penetration depth of the magnetic field if
the
last term is ignored . 
Parameter c is the speed of sound in superconductors. 
The third term is odd under the time-reversal transformation 
and resembles the Chern-Simons term,  
but is not totally antisymmetric about space-time 
indices ($e_{ij}=\varepsilon_{ij}$ in the 2+1 dimensions , and 
$e_{ij}=\varepsilon_{ij3}$ in the 3+1 dimensions). 
Parameters ($m_{\rm B},c,\sigma$) vary  depending on the spatial
dimension. 
For the  A phase, they  are given  in Table I.  


 We study  the characteristic features of T-violating systems based on 
the above action and ignore the
 order-parameter-dependent terms. Order-parameter- dependent 
terms depend on the model used  and  their properties have been 
given in the literature. We  hence ignore them and discuss 
general time-reversal- violating properties derived from eq.(1) in 
the present paper. 

To elucidate  the E-B mixing effects, we first study the $^{3}$He-A- like 
superconductor in a static  electric field. Usually, nothing happens except 
 the screening of the electric field in conventional superconductors,but in 
the present case,  a magnetic field and an electric current 
are induced in the superconductor.  

Suppose there is a superconductor in the $0<x$ region and 
there is a  static background 
electric field ${\rm E}_{x}^{(\rm b)}$ directed along  the x-axis in the 
$x<0$ region. 
We call the $x<0$ region as region ${\rm I}$, and the $0<x$ region as region 
${\rm II}$.  
Thus , there is a boundary at the y-z plane (y-axis) in the 3+1-dimensional 
space-time ( 2+1 dimensional space-time)(Fig.1).


The action in both regions is 
\begin{eqnarray}
& S_{\rm I}=\int d^{\rm D}x [-\frac{1}{4}F_{{\rm I}\mu\nu}F_{\rm I}^{\mu\nu} 
                    -\frac{1}{2}F_{{\rm I}\mu\nu}F_{\rm I}^{({\rm b})\mu\nu} 
                       +\frac{1}{2}{\rm E}_{x}^{({\rm b})2}] &\nonumber\\
&F_{\rm I}^{\mu\nu}=\partial^{\mu} A_{\rm I}^{\nu} - \partial^{\nu} 
A^{\mu}_{\rm I} 
&\nonumber\\
&F_{\rm I}^{({\rm b})10}={\rm E}_{x}^{(\rm b)}, 
{\rm otherwise}~~ 0 &\nonumber\\
& S_{\rm II}=S_{\rm eff}^{\rm (f)} . &\label{20} 
\end{eqnarray}  
From the variational principle, we can derive the equations of motion in  both 
regions under  boundary conditions at $\partial{\rm I} (=-\partial{\rm II})$, 
which is the surface of  region $\rm I$ ( region $\rm II$, but 
opposite orientation). The eqations of motion are written as 
\begin{eqnarray} 
& {\rm in~the~region~I}~;~~\partial_{\mu}F_{\rm I}^{\mu\nu}=0,&\nonumber\\
& {\rm in~the~region~II}~;~~\partial_{\mu}F_{\rm II}^{\mu\nu}=j_{\rm II}^{\nu},
&
\label{30} 
\end{eqnarray}
$j^{\rm II}_{\mu}$ 's are currents and their forms are 
\begin{eqnarray}
&j^{\rm II}_{0}=-\frac{m_{\rm B}^{2}}{c^{2}}A^{\rm II}_{0}
               -\sigma e_{ij}\nabla_{i}A^{\rm II}_{j}
               +\frac{\sigma}{2}A^{\rm II}_{y}\delta(x) &\label{34}\\
&j^{\rm II}_{i}=-{m_{\rm B}^{2}}A^{\rm II}_{i}
               +\sigma e_{ij}\nabla_{j}A^{\rm II}_{0}
               +\frac{\sigma}{2}e_{1i}A^{\rm II}_{0}\delta(x) &\label{35}.
\end{eqnarray}
The first term on  the right hand side  of eq.(\ref{35}) shows 
the London current, 
the second is the Hall current which flows perpendicular to the electric field,
 and the third is an edge current which exists 
only on the boundary surface. 
The latter two terms arise  from the induced T-violating term and its surface 
term at the boundary. 
The boundary conditions are written as   
\begin{eqnarray}
&\int_{\partial {\rm I}} 
d\sigma_{\mu}[-F_{\rm I}^{\mu\nu}-F_{\rm I}^{({\rm b})\mu\nu}+
F_{\rm II}^{\mu\nu}]\delta A_{\nu}&\nonumber\\
&-\frac{\sigma}{2}\int_{\partial {\rm I}}d\sigma_{i}
e^{ij}(A^{\rm II}_{0}\delta A^{\rm II}_{j}-A^{\rm II}_{j}
\delta A^{\rm II}_{0})=0, &\label{40}   
\end{eqnarray}
where $d\sigma_{\mu}$ is a unit area vector of  surface $\partial{\rm I}$.

We solve gauge fields of static configurations, which  have 
only x-dependence  because of the spatial symmetry of the system. 
We assume  that in region I there is a  constant electric field. 
Hence, $A_{\mu}^{\rm I}$ should be constant and  are connected 
 continuously with gauge fields in  region II at the boundary. 
Equations of motion in  region II reduce to  
\begin{eqnarray}
&\frac{d^{2}}{dx^{2}}A^{{\rm II}}_{0}(x)
=\frac{m_{\rm B}^{2}}{c^{2}}A^{\rm II}_{0}(x)+
\sigma \frac{d}{dx}A^{\rm II}_{y}(x)&\nonumber\\
&\frac{d^{2}}{dx^{2}}A^{{\rm II}}_{i}(x)
=m_{\rm B}^{2}A^{\rm II}_{i}(x)+\sigma e_{1i}\frac{d}{dx}A^{\rm II}_{0}(x). &
\label{50}
\end{eqnarray}
$A^{\rm II}_{x}$ and $A^{\rm II}_{z}$ (only in the 3+1 dimensional case) 
satisfy the ordinary London equation, but $A^{\rm II}_{0}$ and 
$A^{\rm II}_{y}$ are influenced by the T-violating term and they mix.
 Since  $c<<1$ (Table I), eigenvalues of the mixing equation are  
\begin{equation}
q_{+} \simeq \frac{m_{\rm B}}{c}~,~q_{-} \simeq m_{\rm B},
\end{equation}
and corresponding eigenvectors are written as 
\begin{equation}
\vec{v}_{+}=
{\rm N}_{+}\left( \begin{array}{c}
 q_{+}^{2} - m_{\rm B}^{2} \\
 - \sigma q_{+}
 \end{array} \right),
\vec{v}_{-}=
{\rm N}_{-}\left( \begin{array}{c}
 -  \sigma q_{-}  \\
 q_{-}^{2} - \frac{m_{\rm B}^{2}}{c^{2}}
 \end{array} \right)
\end{equation}
Both $N_{+}$ and $ N_{-}$ are normalization 
constants for each eigenvector. Using these eigenvalues and eigenvectors, 
$A_{0}$ and $A_{y}$ are written as 
\begin{equation} 
\left( \begin{array}{c} 
A_{0} \\ 
A_{y} 
\end{array} \right)
={\rm C_{0}} \vec{v}_{+} e^{- q_{+} x} + 
 {\rm C}_{y} \vec{v}_{-} e^{- q_{-} x}. 
\end{equation}
Amplitudes $C_{0}$,and $C_{y}$  are decided by boundary 
conditions (\ref{40}) at the surface, $x=0$,  
which  reduce to  
\begin{eqnarray}
&\lim_{x \rightarrow +0}[-\frac{d}{dx}A^{\rm II}_{0}(x)+
\frac{\sigma}{2}A^{\rm II}_{y}(x)]={\rm E}_{x}^{{\rm (b)}}&\nonumber\\
&\lim_{x \rightarrow +0}[\frac{d}{dx}A^{\rm II}_{y}(x)
-\frac{\sigma}{2}A^{\rm II}_{0}(x)]=0. &\label{bc}
\end{eqnarray}

As a result of  E-B mixing a  magnetic field is 
induced in  region II ({\it i.e.}, 
$B^{\rm II}_{z}(x)=\frac{d}{dx}A^{\rm II}_{y}(x)$).   
By taking the leading terms of  the power expansion of c,   
the solution is written as 
\begin{equation} 
B^{\rm II}_{z}(x) \simeq {\rm E}_{x}^{({\rm b})}
\frac{c \sigma}{m_{\rm B}}[ e^{- m_{\rm B} x / c}  
-\frac{1}{2} e^{- m_{\rm B} x}]. 
\label{80}
\end{equation}
According to eq.(\ref{80}), 
discontinuity of the magnetic field arises at 
the boundary, and its magnitude is 
$ {\cal{O}}(\frac{c \sigma}{m_{\rm B}} {\rm E}_{x}^{\rm (b)})$. 
This originates  from an extra term,  
$\lim_{x \rightarrow +0} \frac{\sigma}{2} A_{0} $,  
in the boundary condition for the magnetic field in eqs.(\ref{bc}).    
Spatial dependence of the field is specified by two length scales,  
$m_ {\rm B}^{-1}$ and $(m_{\rm B} / c)^{-1} $, and the field has the extremum  
value before it tends to zero at $x \rightarrow \infty$. Its behavior is 
shown in Fig.2. 
Using the Maxwell equation, we obtain an 
electric current directed along  the y-axis,which is  written as 
\begin{eqnarray}
&j_{y}^{\rm II}(x) = (\vec{\nabla} \times \vec{B}^{\rm II}(x))_{y} &
\label{denryu}\\ 
&\simeq -\sigma {\rm E}_{x}^{(\rm b)}[-e^{- m_{\rm B} x / c} + 
\frac{ c}{2} e^{- m_{\rm B} x} 
- \frac{c}{2 m_{\rm B}} \delta (x)].\nonumber &
\end{eqnarray}
 We are able to derive the same result from eq.(\ref{35}).  
As  mentioned previously , the current consists of three components, the 
London current, the Hall current, and the edge current. 
The Hall current is written as 
\begin{eqnarray}
&j_{y \rm (H)}^{\rm II}(x)=  \sigma \nabla_{x} A_{0}^{\rm II}(x)
\label{halldenryu} &\\
& \simeq - \sigma {\rm E}_{x}^{\rm (b)} [e^{- m_{\rm B} x / c} - 
\frac{1}{2} \frac{c^{3} \sigma^{2}}{m_{\rm B}^{2}} e^{- m_{\rm B} x}].
&\nonumber  
\end{eqnarray}
In general, the Hall current occurs in the  background magnetic field,
but in the present case, 
Hall current is  generated not by the  background magnetic field ,but  by 
the induced magnetic field due to E-B mixing.  
 
At the  cylinder surface boundary , a magnetic moment is induced by the above 
Hall current.  
We assign $a$ as the  radius of 
the cylinder, and it takes a value much larger than the length 
 $1 / m_{\rm B}$. 
In this case, we choose the cylinder coordinates. Corresponding to 
$j_{y}^{\rm II}(x)$, an angular current $j_{\theta}^{\rm II}(r)$ exists and 
they have the  relation   
$j_{\theta}^{\rm II}(r) \simeq - j_{y}^{\rm II}(a - r)$ when $a$ is large. 
The  magnetic moment $M_{z}^{\rm ind}$ induced by $j_{\theta}^{\rm II}$ is 
written as 
\begin{equation}
M_{z}^{\rm ind}=e \int d\theta dr r^2 j_{\theta} \simeq 
  e a^2 \frac{2 \pi c \sigma}{m_{\rm B}} {\rm E}_{x}^{\rm {(b)}}
\label{120}.
\end{equation} 
This induced magnetic moment\cite{Laugh} is proportional to the external 
electric field. 
These results eqs.(\ref{80}),(\ref{denryu}), and (\ref{120}) 
originate   from the T-violating term, because 
they vanish when $\sigma=0$.

Instead of the background electric field E$_{x}^{\rm (b)}$, we consider a 
background magnetic field B$_{z}^{\rm (b)}$ directed along  the $z$-axes. 
Except for  boundary conditions (\ref{bc}), 
this calculation is the same as in  the previous case.  
 We obtain the behavior of the magnetic field in  
region II  as 
\begin{equation}
B^{\rm II}_{z}(x) \simeq {\rm B}_{z}^{({\rm b})}
[-\frac{c \sigma^{2}}{2 m_{\rm B}^{2}}e^{- m_{\rm B} x / c}  
+e^{- m_{\rm B} x}]. 
\label{82}
\end{equation}
This shows the Meissner effect in the A-phase of the p-wave 
superconductor. It is unusual because the magnetic 
field has two damping modes. However  this effect is almost invisible  
because one of the mass eigenmodes $m_{\rm B} / c$ 
is much larger than the other one $m_ {\rm B}$, and the coefficient   
of $e^{- m_{\rm B}x / c}$ is much smaller than that of $e^{- m_{\rm B}x}$. 
The electric current 
\begin{equation}
j_{y}^{\rm II}(x) \simeq {\rm B}_{z}^{\rm (b)}
[\frac{\sigma^{2}}{2 m_{\rm B}} e^{- m_{\rm B} x / c} + 
m_{\rm B} e^{- m_{\rm B}x} 
- \frac{c \sigma^{2}}{4 m_{\rm B}^{2}} \delta(x)]\label{denryu2}
\end{equation}
also occurs in the present  case and  is affected by the T-odd term. 
However  the ordinary supercurrent also occurs in superconductors without 
T-violation 
({\it i.e.,} $\sigma = 0$) under the magnetic field.  
It is diffucult  to observe the 
T-violating effect experimentally in  eq.(\ref{denryu2}) 
compared  with the previous case.  
This  is also the same for the induced magnetic moment.


Finally, we show that the T-odd term in  action (\ref{10}) causes 
T-violating 
light scattering at the surface of the superconductor.  
We consider the  reflection of  polarized light injected from the outside 
of the superconductor along the $x$ direction (Fig. 1) 
and calculate the  rotation of the polarization plane in the reflected wave. 
This effect, which we call  Faraday rotation 
here, has already been  investigated by Wen and Zee\cite{farady1}\cite{farady2}
\cite{2+1C-S6} in a
T-violating superconductor. They showed  that a T-odd term in 
the Landau-Ginzburg action, e.g., $\epsilon_{ij} A_{i} \partial_{0} A_{j}$ 
causes this effect. In this letter, we show that this  effect also occurs 
with the T-odd term in eq.(\ref{10}).     

We study time-dependent solutions of eqs.(\ref{30}). 
We obtain a condition for the gauge fields in the superconductor 
by taking the divergence of both sides of eq.(\ref{30}) 
 (omitting the suffix  ) 
\begin{equation} 
\partial^{\mu} j_{\mu}=0 .\label{condi}  
\end{equation}
This condition is nontrivial in the system with  broken gauge invariance , but 
is trivial in the gauge invariant system. Using this condition, eq.(\ref{30})
for  the superconductor 
become as follows, 
\begin{eqnarray}
&(\partial_{0}^{2} - c^{2} \partial_{x}^{2} + m_{\rm B}^{2})A_{0} 
=-  c^{2} \frac{\sigma}{m_{\rm B}^{2}}(\partial_{0}^{2} 
+ m_{\rm B}^{2})\partial_{x}A_{y}
&\label{innereq21}\\
&(\partial_{0}^{2} - c^{2} \partial_{x}^{2} + m_{\rm B}^{2})A_{x} 
= - c^{2} \frac{\sigma}{m_{\rm B}^{2}}\partial_{0}\partial_{x}^{2}A_{y}
&\label{innereq22}\\
&(\partial_{0}^{2} - \partial_{x}^{2} + m_{\rm B}^{2})A_{y} 
= -\sigma \partial_{x}A_{0}
&\label{innereq23}\\
&(\partial_{0}^{2} - \partial_{x}^{2} + m_{\rm B}^{2})A_{z} = 0. 
&\label{innereq24}
\end{eqnarray} 
According to  eqs.(\ref{innereq21}) and eq.(\ref{innereq23}), 
$A_{0}$ and $A_{y}$  mix,  
and their propagation is specified by two 
momentum eigenmodes  $ q_{+} \simeq 
\frac{(k_{0}^{2} - m_{\rm B}^{2})^{1/2}}{c} $ and 
$ q_{-} \simeq (k_{0}^{2} - m_{\rm B}^{2})^{1/2} $ 
,when $c<<1$, where $k_{0}$ is 
the  frequency of  light.   
Equation (\ref{innereq22})shows that  $A_{x}$ is caused by 
$A_{y}$ which would be considered 
as a source of  force vibration.  
$A_{z}$ does not mix with the others. 
Condition (\ref{condi}) suggests that the solution of eq.(\ref{innereq22}) 
does not have the freedom to add  kernel, 
which is the same solution as when the r.h.s. of 
eq.(\ref{innereq22}) is zero.  
The gauge fields in  region II 
should be written as 
\begin{eqnarray}
& A_{\mu} = c_{\mu +} e^{-i k_{0} t + i q_{+} x} + 
c_{\mu -} e^{- i k_{0} t + i q_{-} x}~;~\mu=0,1,2 &\nonumber\\
& A_{z}=c_{z} e^{- i k_{0} t + i \sqrt{k_{0}^2 - m_{\rm B}^{2}}}. 
\label{innergauge} 
\end{eqnarray}
The coefficients $c_{\mu+,-}$,and $c_{z}$ will be chosen to satisfy 
the equations of motion,  condition (\ref{condi}), and the boundary conditions.
The gauge fields in  region I are written as 
\begin{equation}
A_{\mu}=a_{\mu} e^{-i k_{0}(t - x)} + b_{\mu} e^{-i k_{0}(t + x)}. 
\label{aout}
\end{equation} 
The first term on  the r.h.s. of eq.(\ref{aout}) indicates 
the incident wave and the second indicates the  reflected wave. 
Boundary conditions at $x=0$ require that 
the gauge fields connect continuously 
and that their derivatives obey eq.(\ref{40}). 
In this case, there are no background fields in the region I 
({\it i.e.}, ${\rm E}^{\rm (b)}={\rm B}^{\rm (b)}=0$). 
Using the boundary conditions, we determine the ratios of  $c_{\mu +,-},c_{z}$ 
and $b_{\mu}$. 
We set the linearly polarized  incident wave along the  y-direction 
({\it i.e.,} $a_{y}=1,a_{0}=a_{x}=a_{z}=0$).  
$b_{\mu}$ for $k_{0}<<m_{\rm B}$ are written as 
\begin{equation}
b_{0} \simeq  \frac{2 i k_{0} c^2 \sigma}{m_{\rm B}^{2}},
\end{equation}
and  for $k_{0} \simeq m_{\rm B}$, are written as
\begin{equation}
 b_{0} \simeq \frac{4 k_{0}}{\sigma},
       {c \sigma (m_{\rm B}^{2} - k_{0}^{2})^{1/2}}, 
\label{refm}
\end{equation}
In both cases, $b_{z}=0$. The effect of the T-violation is seen in 
$b_{0}$ and $b_{x}$ and the rotation of polarization occurs. 
When $k_{0}\simeq m_{\rm B}$ a resonance occurs, 
but near the resonance, the  damping effect of the vibration 
that is neglected here should be taken into account. 
Furthermore, we should note that 
light absorption of the superconductor occurs in the region of energy 
larger than the gap energy $|\Delta|$ because  quasi-particle 
production occurs in such an energy region. 
Therefore  in the case of $|\Delta|<<m_{\rm B}$, the resonance is not  
 detectable. However  in some superconductors the  
parameters would satisfy $m_{\rm B} < |\Delta| $. 
The energy dependence of the absolute values of 
these amplitudes are given in Fig.3  
in the region $0<k_{0}<m_{\rm B}$ for 
heavy-fermion superconductors and high-T$_{\rm c}$ superconductors
\cite{Sup1,Sup2}\cite{Sr2RuO41,Sr2RuO42}. T-violating amplitude has a 
significant  magnitude.   

In this letter, we showed  some  
unusual electromagnetic phenomena in the A-phase of a p-wave superconductor.
This system violates T spontaneously, 
so it has an induced Chern-Simons-like T-violating term 
in the Landau-Ginzburg effective action. 
Owing to this term, mixing between the electric and  magnetic fields 
 occurs. In the system with  an external electric field, a magnetic 
field and an  
electric current are induced. The current 
includes a Hall current. This  suggests that a  Hall effect occurs  
without the external magnetic field . The current gives the magnetic moment 
of the system. In the system with  a magnetic field, an unusual Meissner 
effect occurs and the magnetic field in the superconductor obeys two damping 
modes.   
We have shown also that T-violating light scattering, such as Faraday rotation 
,occurs at the surface of the p-wave superconductor A-phase.  
Such  effects  that were  studied here  are generated by the 
T-odd Chern-Simons-like term. They should show chacteristic features of 
T-violating systems and may also occur for other T-violating 
superconductors.


The authors are grateful to Dr. N. Maeda for useful discussions.
This work was partially supported by a  special Grant-in-Aid for the Promotion 
of Education and Science in Hokkaido University provided by the Ministry 
of Education, Science, Sports, and Culture,  Grants-in-Aid for Scientific 
Research (07640522),  for Scientific Research on Priority area
(Physics of CP violation), and  for International Science 
Research (Joint Research 07044048) from the Ministry of Education, Science, 
Sports and Culture, Japan.   



\newpage
\vspace{0.5in}

\begin{center}
{\bf Captions}
\end{center}

Table.I. Parameters in A phase. $m_{\rm e},\rho$, and 
$\epsilon_{\rm F}$ are mass, 
number density, and the Fermi energy of electrons. 
\vspace{0.2in}

Fig.1. This Figure shows the situation of our calculation. 
there is a background electric  
field E$_{x}^{(\rm b)}$ directed to the x-axis in the region I, 
and the $^{3}$He-A like superconductor (represented by shadow lines) 
is in the region II. 
\vspace{0.2in}

Fig.2. The induced magnetic field $B_{z}^{\rm II}(x)$ in the 
superconductor under a background 
electric field ${\rm E}_{x}^{(b)}$ directed to x-axis.
\vspace{0.2in}

Fig.3. Energy dependence of the logalizm of 
$|b_{0}|,|b_{x}|$, and $|b_{y}|$ when $a_{y}=1$ and others are 0. 
Parameters $c, \sigma, m_{\rm B}$ 
are chosen as typical values in High-T$_{\rm c}$ superconductors (solid lines) 
and heavy fermion superconductors (dashed lines).

\vspace{0.5in}
Table.I
\begin{center}
\begin{tabular}{ccccc} \hline 
parameters & & 2+1 dimension & & 3+1 dimension \\ \hline 
$m_{\rm B}^{2}$ & & $\frac{\rho e^{2}}{m_{\rm e}}$ & & 
$\frac{\rho e^{2}}{m_{\rm e}}$ \\ \hline
$c^{2}$ & & $\frac{\rho}{m_{\rm e}^{2}}$ & & 
$\frac{2 \epsilon_{\rm F}}{3 m_{\rm e}}$ \\ \hline
$\sigma$ & & $\frac{e^{2}}{4 \pi}$ & &  
$\frac{\rho e^{2}}{8 m_{\rm e} \epsilon_{\rm F}}$  \\ \hline \label{parameter}
\end{tabular}
\end{center}

\end{document}